\documentclass[twocolumn]{aastex631}
\usepackage{bm}	
\usepackage{graphicx}	% Including figure files
\usepackage{amsmath}	% Advanced maths commands
\usepackage[normalem]{ulem}

\shorttitle{Constraints on Primordial Magnetic Fields from High Redshift Stellar Mass Density}
\shortauthors{Qile Zhang et al.}

\begin{document}

\title{Constraints on Primordial Magnetic Fields from High Redshift Stellar Mass Density}

\author{Qile Zhang}
\affiliation{Department of Astronomy, Beijing Normal University, Beijing 100875, China}
\author{Shang Li}
\affiliation{Department of Astronomy, Beijing Normal University, Beijing 100875, China}
\author{Xiu-Hui Tan}
\affiliation{Department of Astronomy, Beijing Normal University, Beijing 100875, China}
\author{Jun-Qing Xia}\thanks{xiajq@bnu.edu.cn}
\affiliation{Department of Astronomy, Beijing Normal University, Beijing 100875, China}
\affiliation{Institute for Frontiers in Astronomy and Astrophysics, Beijing Normal University, Beijing 100875, China}

\begin{abstract}

Primordial magnetic fields (PMFs) play a pivotal role in influencing small-scale fluctuations within the primordial density field, thereby enhancing the matter power spectrum within the context of the $\Lambda$CDM model at small scales. These amplified fluctuations accelerate the early formation of galactic halos and stars, which can be observed through advanced high-redshift observational techniques. Therefore, Stellar Mass Density (SMD) observations, which provide significant opportunities for detailed studies of galaxies at small scales and high redshifts, offer a novel perspective on small-scale cosmic phenomena and constrain the characteristics of PMFs. In this study, we compile 14 SMD data points at redshifts $z > 6$ and derive stringent constraints on the parameters of PMFs, which include the amplitude of the magnetic field at a characteristic scale of $\lambda=1\,{\rm Mpc}$, denoted as $B_0$, and the spectral index of the magnetic field power spectrum, $n_{\rm B}$. At 95\% confidence level, we establish upper limits of $B_0 < 4.44$ nG and $n_{\rm B} < -2.24$, along with a star formation efficiency of approximately $f_*^0 \sim 0.1$. If we fix $n_{\rm B}$ at specific values, such as $-2.85$, $-2.9$, and $-2.95$, the 95\% upper limits for the amplitude of the magnetic field can be constrained to 1.33 nG, 2.21 nG, and 3.90 nG, respectively. Finally, we attempt to interpret recent early observations provided by James Webb Space Telescope (JWST) using the theory of PMFs, and find that by selecting appropriate PMF parameters, it is possible to explain these results without significantly increasing the star formation efficiency.

\end{abstract}

\keywords{cosmology, Primordial Magnetic Field, high redshift, star formation efficiency -- small-scale structure of Universe}
%%%%%%%%%%%%%%%%%%%%%%%%%%%%%%%%%%%%%%%%%%%%%%%%%%
\section{Introduction}\label{section:1}

Recent advancements in cosmological probes, including the Cosmic Microwave Background (CMB), Baryon Acoustic Oscillations (BAO), Supernovae (SN), and Redshift-Space Distortions (RSD), have facilitated the acquisition of extensive information on the large to medium scales of the universe. These developments enhance our understanding of cosmological evolution and, within the framework of the $\Lambda$CDM model, enable precise constraints on numerous cosmological parameters. As galaxy surveys advance, our capacity to detect increasingly smaller scales has improved \citep{2019MNRAS.489.2247C}. This progress necessitates the use of appropriate methodologies to extract information from these smaller scales, where a wealth of non-standard cosmological phenomena predominantly reside.

Magnetic fields, pervasive across various cosmic environments—from planets to galaxy clusters—exhibit strengths ranging from approximately 1 G to $10^{-6}$ G. These fields are coherent over scales up to tens of kiloparsecs within and around high-redshift galaxies and in the interstellar medium{\citep{1996ARA&A..34..155B,doi:10.1126/science.1154923,doi:10.1142/S0218271804005080}.
The strongest magnetic fields discovered in the universe are found in neutron stars, particularly those in X-ray binaries, with strengths ranging from approximately $10^{8}$ to several $10^{13}$ G \citep{2015SSRv..191..293R}.
The generation of these magnetic seeds, whether through mechanisms like the Biermann battery \citep{1950ZNatA...5...65B} or the influence of the first stars \citep{2011ApJ...741...93D,2010ApJ...721L.134S}, and their amplification via dynamo processes \citep{PhysRevLett.107.114504,PhysRevE.85.026303}, is considered crucial. 
If a field in the intergalactic medium (IGM) fills the vast, empty regions of space (voids), it would be challenging to explain this solely through astrophysical processes that occur in the late universe. This difficulty suggests that such a field might have a primordial origin \citep{Adi_2023,10.1111/j.1365-2966.2006.10474.x,2001ApJ...556..619F,Subramanian_2016}. 
Magneto genesis during inflation or early phase transitions is particularly important, which can be categorized into: (1) generation during phase transitions, producing very blue or violet spectra which have $n_B\sim 2$  \citep{PhysRevLett.51.1488,GRASSO1998258,GRASSO2001163}, and (2) inflationary generation, resulting in nearly scale-invariant spectra which have $n_B\sim -3$ \citep{1992ApJ...391L...1R,PhysRevD.37.2743, PhysRevLett.123.021301,Durrer_2013}. 
For instance, PMFs generated during inflation, scalar fields ($\phi$) like the inflaton and the dilaton such as $f^2(\phi) \sim \mathrm{e}^{\alpha \phi}$, which can be expected that $B_0 \sim 10^{-9}$ to $10^{-65}$ G, for $\alpha \sim 20-0$ \citep{1992ApJ...391L...1R}.
PMFs generated during phase transitions, such as Higgs field gradients which lead to a comoving field $B_0 \sim 7 \times 10^{-8} \mathrm{G}$ \citep{VACHASPATI1991258, GRASSO1998258,Subramanian_2016}.
These PMFs primarily influence small-scale matter density perturbations, which is the main focus of our study.

In recent years, several constraints on the parameters of PMFs have been derived from various astronomical observations. 
For instance, vortical modes generated by PMFs can dominate the B-mode power spectrum on small scales and affect the Silk damping scale. Consequently, the CMB provides the stringent limits on PMFs due to their significant contribution to the B-mode polarization power spectrum. \citet{PhysRevD.95.063506} utilized the $Planck$ CMB temperature and polarization data to constrain the magnetic amplitude, obtaining an upper limit of $B_{1 \mathrm{Mpc}} < 3.3$ nG at the 95\% confidence level. When using future CMB observational data, such as those from the LiteBIRD satellite, the constraint is expected to improve, with a marginalized limit of $B_{1 \mathrm{Mpc}}^{\operatorname{marg}} < 2.2$ nG at the 95\% confidence level \citep{paoletti2024litebird}. By incorporating data at large multipoles from the South Pole Telescope (SPT) and marginalizing over the spectral index $n_B$, this upper limit is further refined to 1.5 nG, with $n_B$ remaining unconstrained.

Besides CMB observations, various astrophysical observations can also provide constraints on PMFs. For instance, \citet{sanati2024dwarf} used magneto-hydrodynamics methods to investigate the impact of magnetic fields on the evolution of dwarf galaxies. They found that dwarf galaxies formed in a $\Lambda \mathrm{CDM}$+PMF scenario with $B_0 = 0.05$ and $0.10$ nG align more closely with observed scaling relations in the Local Group. 
Additionally, \citet{Adi_2023} employed the voxel intensity distribution (VID) derived from the matter power spectrum, using upcoming Line-Intensity Mapping (LIM) surveys targeting carbon monoxide (CO) emission from star-forming galaxies at high redshifts. This approach can constrain the parameters of PMFs to values as low as $B_{1 \mathrm{Mpc}} \sim 0.006-1$ nG.
Most of the studies mentioned above focus on constraining the upper limit of $B_0$. However, it is worth noting that, based on blazar emissions, it is suggested that the voids in the intergalactic medium might host a weak magnetic field of about $10^{-16}$ G, coherent on Mpc scales. This indicates a lower limit for PMFs such that $B_0 \geq 3 \times 10^{-16}$ G \citet{2010Sci...328...73N,2011PhR...505....1K}.

With the advancement of telescopes and satellites, extensive surveys have yielded vast amounts of data. Despite this abundance, small-scale data on galaxies and clusters remains essential, especially due to its dependence on the intricate physics of star formation. The Stellar Mass Function (SMF) and Stellar Mass Density (SMD) of galaxies provide significant opportunities for detailed studies of galaxies at small scales and high redshifts \citep{Bhatawdekar_2019,2020ApJ...893...60K,2021ApJ...922...29S}. The presence of PMFs significantly enhances the matter power spectrum at small scales. Consequently, the SMD, derived from the matter power spectrum, is expected to exhibit differences. Therefore, we undertake an analysis utilizing current SMD observations to constrain the parameters of PMFs.

The paper is structured as follows: In Section \ref{section:2}, we briefly introduce the theory of PMFs and their influence on the matter power spectrum. Section \ref{section:3} provides a concise overview of star formation and galaxy formation theories, illustrating their relevance to PMFs. Section \ref{section:4} presents the numerical method and SMD data employed in this study. Our results regarding the constraints on PMFs are presented and discussed in Section \ref{section:5}. Finally, we conclude our work in Section \ref{section:6}.

This work assumes a flat universe with the $Planck$ 2018 best-fit parameters $h=0.6736$, $\Omega_{\mathrm{m}}  = 0.3153$, $\Omega_{\mathrm{b}}  = 0.0493$ \citep{2020A&A...641A...6P}.
%%%%%%%%%%%%%%%%%%%%%%%%%%%%%%%%%%%%%%%%%%%%%%%%%%
\section{Primordial magnetic field}\label{section:2}
The PMFs can be divided into two categories, one is helical field, the other is non-helical field. 
The conservation of magnetic helicity plays an important role, leading to a larger coherence scale compared to a non-helical field \citep{PhysRevE.64.056405}.
More specifically, using the angular power spectra and the Planck likelihood, the 95\% confidence level constraints are $B_0 < 5.6 $ nG for helical fields.
For those models describing the generation of magnetic fields during the electroweak phase transition (EWPT), it is estimated that the strength of helical magnetic fields on a scale of 10 kpc is approximately $B_{10 \mathrm{kpc,helical}}\sim 10^{-11}$ G \citep{2021RPPh...84g4901V}.

To briefly consider the effects of PMF-induced power spectrum,
we limit ourselves to considering a non-helical magnetic field because the density perturbations, and consequently the matter power spectrum, are not influenced by the presence of magnetic helicity \citep{Kahniashvili_2013}. For inflationary generation, magnetic fields are also non-helical \citep{2011PhR...505....1K,Durrer_2013}.

We assume that non-helical magnetic fields $\boldsymbol{B}$ are isotropic and homogeneous, so we can express the two-point correlation function of the magnetic field in Fourier space as follows:
\begin{equation}
\langle\hat{B}_i(\boldsymbol{k}) \hat{B}_j^*(\boldsymbol{k}^{\prime})\rangle = (2 \pi)^3 \delta(\boldsymbol{k}-\boldsymbol{k}^{\prime}) \frac{P_{i j}(\boldsymbol{k})}{2} P_B(k)~,
\end{equation}
where $\mathrm{P}_{i j}(\boldsymbol{k})=\delta_{i j}-\frac{k_i k_j}{k^2}$ and $P_B(k, t)$ represents the PMFs power spectrum. In most cases, we consider the magnetic power spectrum to follow a power-law form:
\begin{equation}
P_B(k, t) = A_B(t) k^{n_{B}},
\end{equation}
where $n_{\rm B}$ denotes the spectral index, which depends on the mechanism generating the PMFs.
%%%%%%%%%%%%%%%%%%%%%%%%%%%%%%%%%%%%%%%%%%%%%%%%%%
\subsection{Magnetic fields damping}

As the temperature decreases post-Big Bang, reaching approximately $10^4$ K at $z \sim 1100$, free electrons and protons combined to form neutral hydrogen \citep{2020A&A...641A...6P}. Prior to this epoch, any hydrogen produced was swiftly ionized by energetic photons, and magnetic field perturbations were predominantly frozen in the baryon-photon plasma, dissipating on scales smaller than the radiation diffusion length. Consequently, the magnetic field modes $\tilde{\boldsymbol{B}}(\boldsymbol{k}, t)$ acquire a damping factor approximated by $\exp{[-k^2 / k_A^2]}$ to account for Alfvén Damping \citep{Adi_2023,PhysRevD.86.043510,10.1093/mnrasl/slx195,10.1093/pasj/psac015,PhysRevD.81.043517}, where the Alfvén wave number $k_A$ is determined by
\begin{equation}
\frac{1}{k_A^2} = \int_0^{t_{\mathrm{rec}}} \frac{v_A^2 \tau_c}{a^2(t)} dt~,
\end{equation}
where $1 / \tau_c = c \cdot n_e \cdot \sigma_T$ represents the Thomson scattering rate.

On the other hand, we also need to consider the Alfvén velocity, which is determined from the pressure and energy density of the plasma and the magnetic field, and can be expressed as:
\begin{equation}
v_A = \frac{B_{\lambda_A}(t)}{\sqrt{\mu_0[\rho(t)+p(t)]}} = \frac{B_{\lambda_A}(t)}{\sqrt{\mu_0[\bar{\rho}_b(t)+\frac{4}{3} \bar{\rho}_r(t)]}}~,
\end{equation}
whereas
\begin{equation}
\frac{1}{k_A^{n_B+5}} = \frac{B_0^2 \cdot (1 \text{ Mpc})^{n_B+3}}{\mu_0 \cdot (2 \pi)^{n_B+3}}\cdot \int_{z_{\text {rec }}}^{\infty} \frac{(1+z)^5 d z}{n_e \sigma_{\mathrm{T}}H \cdot[\bar{\rho}_b(z)+\frac{4}{3} \bar{\rho}_r(z)]}~.
\end{equation}
Consequently, the damped PMFs power spectrum becomes:
\begin{equation}
P_B(k, t) = A_B(t) k^{n_B} e^{-2 k^2 / k_A^2}~.
\end{equation}
We need to characterize the magnetic field in terms of its physical strength smoothed over a characteristic scale $\lambda$, denoted as:
\begin{equation}
\begin{aligned}
B_{0,\lambda}^2(t) & =\int_0^{\infty} \frac{d k k^2}{2 \pi^2} P_B(k, t) e^{-k^2 \lambda^2} \\
& =\frac{A_B(t)}{(2 \pi)^2} \frac{\Gamma[(n_B+3) / 2]}{\lambda^{n_B+3}}~.
\end{aligned}
\end{equation}
In this paper, we use $B_0$ defined as $\lambda=1$ Mpc. The amplitude of the PMFs power spectrum is given by:
\begin{equation}
A_B = \frac{(2 \pi)^{n_B+5} B_0^2}{2 \Gamma(\frac{n_B}{2}+\frac{3}{2})}~.
\end{equation}
%%%%%%%%%%%%%%%%%%%%%%%%%%%%%%%%%%%%%%%%%%%%%%%%%%
\subsection{Magnetic Jeans scale }
Prior to the recombination epoch, the electrically conducting fluid experiences a Lorentz pressure, serving as an additional source of matter density fluctuations due to the presence of magnetic fields. On sufficiently small scales, the magnetic pressure gradients counteract the gravitational pulling force, leading to the establishment of a magnetic Jeans scale beyond which the magnetically induced perturbations cease to grow. Consequently, the total matter power spectrum exists only for $k < k_J = 2 \pi/\lambda_J$, where $\lambda_J$ takes the form:
\begin{equation}
\lambda_J=\frac{2 \pi}{k_J}=\left[\frac{16 \pi}{25} \frac{B_{0,\lambda}^2}{\mu_0 G \bar{\rho}_{m, 0} \bar{\rho}_{b, 0}} \lambda^{3+n_B}\right]^{\frac{1}{5+n_B}}~,
\end{equation}
where $\mu_0$ represents the magnetic permeability of vacuum, $G$ denotes Newton’s constant, and $\bar{\rho}_{m, 0}$ and $\bar{\rho}_{b, 0}$ stand for today’s background matter and baryon energy densities, respectively. Note that magnetic Jeans length is independent of time, since the magnetic field evolves as $\mathbf{B}(\mathbf{x}, t)=\mathbf{B}(\mathbf{x}) / a^2(t)$ \citep{Fedeli_2012,1996ApJ...468...28K,1978ApJ...224..337W}.
%%%%%%%%%%%%%%%%%%%%%%%%%%%%%%%%%%%%%%%%%%%%%%%%%%
\subsection{Impact on the matter power spectrum}
The total matter power spectrum of inflationary and PMF-induced perturbations, as given by \citet{PhysRevD.108.023521}, is expressed as:
\begin{equation}\label{PMF_Pk}
P_m(k, t)=D_{+}^2(t) P^{\rm{lin}}(k)+M^2(t) \Pi(k)~,
\end{equation}
where $D_{+}(t)$ represents the cosmological growth factor, normalized such that $D_{+}(t=t_0)\equiv 1$ at the present day. Here, $M^2(t) \Pi(k)$ denotes the PMF-induced matter power spectrum, with the magnetic growth factor $M(t)$ characterizing the temporal evolution of PMF-induced matter perturbations. This evolution is determined by:
\begin{equation}
\ddot{M}(t)+2 H(t) \dot{M}(t)-4 \pi G \bar{\rho}_{m, 0} \frac{M(t)}{a^3(t)}=\frac{1}{a^3(t)}~,
\end{equation}
where we assume the initial conditions $M(t_{\mathrm{rec}})=\dot{M}(t_{\mathrm{rec}}) \equiv 0$ at recombination ($z \sim 1100$). Therefore, we obtain:
\begin{equation}
M(t)=t_0^2\left[\frac{9}{10} \cdot \frac{1+z_{\mathrm{rec}}}{1+z}+\frac{3}{5}\left(\frac{1+z}{1+z_{\mathrm{rec}}}\right)^{\frac{3}{2}}-\frac{3}{2}\right]~,
\end{equation}
where $t_0=({6 \pi G \bar{\rho}_{m, 0}})^{-1/2}$.

According to the formalism developed in \citet{PhysRevD.108.023521}, the $\Pi(k)$ power spectrum due to the PMFs can be evaluated using:
\begin{equation}
\Pi(k)=\frac{f_b^2}{(\mu_0 \bar{\rho}_{b, 0})^2} \int d q \int d \mu \frac{P_{B, 0}(q) P_{B, 0}[\alpha(\boldsymbol{k}, \boldsymbol{q})]}{\alpha^2(\boldsymbol{k}, \boldsymbol{q})} \mathcal{F}(\boldsymbol{k}, \boldsymbol{q})~,
\end{equation}
where $\alpha=|\boldsymbol{k}-\boldsymbol{q}|$, $\mu=\hat{\boldsymbol{k}} \cdot \hat{\boldsymbol{q}}$, and
\begin{equation}
\mathcal{F}(\boldsymbol{k}, \boldsymbol{q})=2 k^5 q^3 \mu+k^4 q^4(1-5 \mu^2)+2 k^3 q^5 \mu^3~.
\end{equation}
In practice, we employ the formalism developed in
\begin{equation}\label{PiK}
\begin{aligned}
\Pi(k)= & (\frac{\alpha k}{4 \pi})^2 \int_0^{\infty} k_1^2 d k_1 \int_{-1}^1 d \mu \\
& \times P_B(k_1) P_B(\sqrt{k^2+k_1^2-2 k k_1 \mu}) \\
& \times[k^2+(k^2-2 k k_1 \mu) \mu^2]~,
\end{aligned}
\end{equation}
where $\alpha \equiv f_b /(\mu_0 \bar{\rho}_{b, 0})$. This expression can avoid situations where the denominator of the integrand approaches zero. We generated our results by incorporating PMF-induced matter perturbations into the publicly available software developed by \citet{2018ApJS..239...35D}.

%%%%%%%%%%%%%%%%%%%%%%%%%%%%%%%%%%%%%%%%%%%%%%%%%%
\section{Model}\label{section:3}

To analyze the impact of PMFs on the SMD, we need to calculate the halo mass function. We consider the halo mass function obtained from the ellipsoidal collapse model \citep{Press-Schechter,Sheth-Tormen-1999,Sheth-Tormen-2001,Cooray-2002}, which can be expressed as:
\begin{equation}\label{halomass}
n(M) d M=\frac{\bar{\rho}}{M} f(\nu) d \nu~,
\end{equation}
where $M$ is the halo mass, $\bar{\rho}$ is the mean comoving matter density, and $\nu(M) = \delta_{\mathrm{c}} / \sigma(M)$. Here, we adopt the threshold overdensity $\delta_{\mathrm{c}}=1.68$. The function $f(\nu)$ is given by:
\begin{equation}
f(\nu)=A[1+(q \nu^2)^{-p}] e^{-q \nu^2 / 2}~,
\end{equation}
where $p = 0.3$, $q = 0.707$, and $A = 0.2161$ are parameters derived from simulations.

The variance of the linear matter overdensity field smoothed on a comoving scale $R$, denoted as $\sigma^2(R)$, can be calculated as:
\begin{equation}
\sigma^2(R)=\int_0^{\infty} 4 \pi(\frac{k}{2 \pi})^3 P^{\operatorname{lin}}(k) W^2(k R) d \ln k~,
\end{equation}
where $W(x)=(3 / x^3)(\sin x-x\cos x)$ is the Fourier transform of a spherical top-hat filter window function. Here, $P^{\operatorname{lin}}(k)$ (referred to as $P(k)$ later) represents the linear matter power spectrum, which can account for both CDM and PMF linear power spectra in this study.

After obtaining the halo mass function, we can estimate the comoving cumulative halo mass density with halo mass greater than $M$ using:
\begin{equation}\label{rho}
\rho(>M, z)=\int_M^{\infty} d M^{\prime} M^{\prime} n(M^{\prime}, z)~.
\end{equation}
Considering the relation between the stellar mass $M_\ast$ and the halo mass, $M_*=(\Omega_{\mathrm{b}} / \Omega_{\mathrm{m}}) f_* M=\epsilon M$, the cumulative stellar mass density with stellar mass greater than $M_\ast$ can be expressed as:
\begin{equation}\label{rho_star}
\rho_*(>M_*, z)=\epsilon \rho(>M_* / \epsilon, z)~.
\end{equation}

The star formation efficiency (SFE) $f_*$, indicating the fraction of baryons that can convert to stars, is considered as a mass-dependent quantity. Following the approach described in \citet{2017MNRAS.464.1365M}, we assume the star formation efficiency follows a double power law given by:
\begin{equation}
f_*(M)=\frac{2 f_*^0}{(\frac{M}{M_{\mathrm{p}}})^{\gamma_{\mathrm{low}}}+(\frac{M}{M_{\mathrm{p}}})^{\gamma_{\mathrm{high}}}}~,
\end{equation}
where $f_*^0$ is the star formation efficiency at its peak mass $M_{\mathrm{p}}$, and $\gamma_{\mathrm{low}}$ and $\gamma_{\mathrm{high}}$ describe the power-law indices at low and high masses. For this study, we adopt $M_{\mathrm{p}}=2.8 \times 10^{11} M_{\odot}$, $\gamma_{\mathrm{low}}=0.49$, and $\gamma_{\mathrm{high}}=-0.61$ for $z \gtrsim 6$.
%%%%%%%%%%%%%%%%%%%%%%%%%%%%%%%%%%%%%%%%%%%%%%%%%%
\begin{table}
\centering
\caption{High redshift SMD data used in our study.}
\renewcommand{\arraystretch}{1.5}
\scriptsize
\tabcolsep=0.16cm
\begin{tabular}{c|c|c|c|c} 
\hline
$z$  & $\rho_*(M > 10^8M_{\sun} )$ & $\sigma_+$ & $\sigma_-$ & Reference\\
& $[\log _{10} M_{\odot} \mathrm{Mpc}^{-3}]$ & & & \\
\hline
        5.9   & 6.99 & 0.09 & 0.09 &\citet{2011ApJ735L34G}\\
        6     & 6.48 & 0.31 & 0.18 &\citet{2016ApJ8255S}\\
        6     & 6.59 & 0.05 & 0.07 &\citet{Grazian_2015}\\
        6     & 6.68 & 0.09 & 0.11 &\citet{2021ApJ...922...29S}\\
        6     & 6.76 & 0.11 & 0.12 &\citet{10.1093/mnras/stu1622}\\
        6     & 6.79 & 0.13 & 0.12 &\citet{Bhatawdekar_2019}\\
        6.8   & 6.84 & 0.14 & 0.17 &\citet{2011ApJ735L34G}\\
        7     & 6.04 & 0.37 & 0.16 &\citet{Grazian_2015}\\
        7     & 6.19 & 0.62 & 0.40 &\citet{2016ApJ8255S}\\
        7     & 6.26 & 0.13 & 0.17 &\citet{2021ApJ...922...29S}\\
        7     & 6.54 & 0.52 & 0.55 &\citet{Bhatawdekar_2019}\\
        7     & 6.64 & 0.56 & 0.89 &\citet{10.1093/mnras/stu1622}\\
        8     & 5.50 & 0.83 & 0.81 &\citet{2016ApJ8255S}\\
        8     & 5.69 & 0.83 & 0.81 &\citet{Bhatawdekar_2019}\\
        8     & 5.73 & 0.21 & 0.33 &\citet{2021ApJ...922...29S}\\
        8     & 6.16 & 0.56 & 0.48 &\citet{2020ApJ...893...60K}\\
        9     & 5.61 & 0.92 & 0.90 &\citet{Bhatawdekar_2019}\\
        9     & 6.32 & 0.76 & 0.64 &\citet{2020ApJ...893...60K}\\
        9     & 4.89 & 0.25 & 0.29 &\citet{2021ApJ...922...29S}\\
        10    & 3.68 & 0.52 & 0.79 &\citet{2021ApJ...922...29S}\\

\hline
\end{tabular}
\label{table_20points}
\end{table}
%%%%%%%%%%%%%%%%%%%%%%%%%%%%%%%%%%%%%%%%%%%%%%%%%%
\section{Method \& Data}\label{section:4}

In our study, we perform an analysis by using the Markov Chain Monte Carlo (MCMC) method. We utilize the Python packages $emcee$\footnote{\url{https://emcee.readthedocs.io}} \citep{2013PASP..125..306F} and $cobaya$\footnote{\url{https://cobaya.readthedocs.io}} \citep{Torrado_2021}. $emcee$, affectionately known as the MCMC Hammer, is widely used for MCMC sampling, while $cobaya$ offers a versatile framework for Bayesian analysis of cosmological models. To visualize the results and derive contours, we employ $getdist$\footnote{\url{https://getdist.readthedocs.io}}, a package designed specifically for analyzing MCMC samples and generating posterior distributions.

Here, we fix the basic cosmological parameters as the best values from the Planck measurement and only vary three primary free parameters: the amplitude $B_0$, the spectrum index $n_{\rm B}$, and the SFE parameter $f_*^0$. Furthermore, to incorporate the impact of galaxy theory on our findings, we also include an additional nuisance parameter $\rm log_{10} a$, denoted as the scatter factor (in dex). We impose a Gaussian prior $\mathcal N(0, 0.15)$ on $\rm log_{10} a$ following the methodology outlined in \citet{2018MNRAS.477.1822M}.

In our study, we compile 20 SMD data points at $z \gtrsim 6$ from various sources \citep{
2010.18118.x,2011ApJ735L34G,10.1093/mnras/stu1622,Grazian_2015,2016ApJ8255S,Bhatawdekar_2019,2020ApJ...893...60K,2021ApJ...922...29S} and present them in Table \ref{table_20points}. Each data point is assumed to follow a Gaussian distribution to establish the individual likelihood, facilitating the derivation of the cumulative stellar mass density at a given redshift:
\begin{equation}\label{likelihood}
\mathcal{L}(\rho_{\ast,i},z_i)=\frac{1}{\sqrt{2\pi\sigma_i^2}}\exp{\left[-\frac{(\rho_{\ast,i}-\rho^{\rm th}_\ast(z_i))^2}{2\sigma_i^2}\right]}~.
\end{equation}
It is important to note that for SMD data points with asymmetric error bars, we adopt the larger error to calculate the likelihood in Eq.(\ref{likelihood}) for the MCMC analysis.
%%%%%%%%%%%%%%%%%%%%%%%%%%%%%%%%%%%%%%%%%%%%%%%%%%

\section{Results and Discussion}\label{section:5}
\subsection{Impact of PMFs on $P(k)$}

As illustrated in \citet{PhysRevD.108.023521}, PMFs predominantly amplify the matter power spectrum at small scales. Furthermore, \citet{PhysRevLett.132.061002} illustrated the importance of enhancements in $P(k)$, such as bumps, in addressing phenomena like the excess of ultramassive galaxy candidates. They emphasize that the parameter $f_*$ plays a crucial role in determining both the position and magnitude of such enhancements. 

\begin{figure}
\centering
\includegraphics[width=\columnwidth]{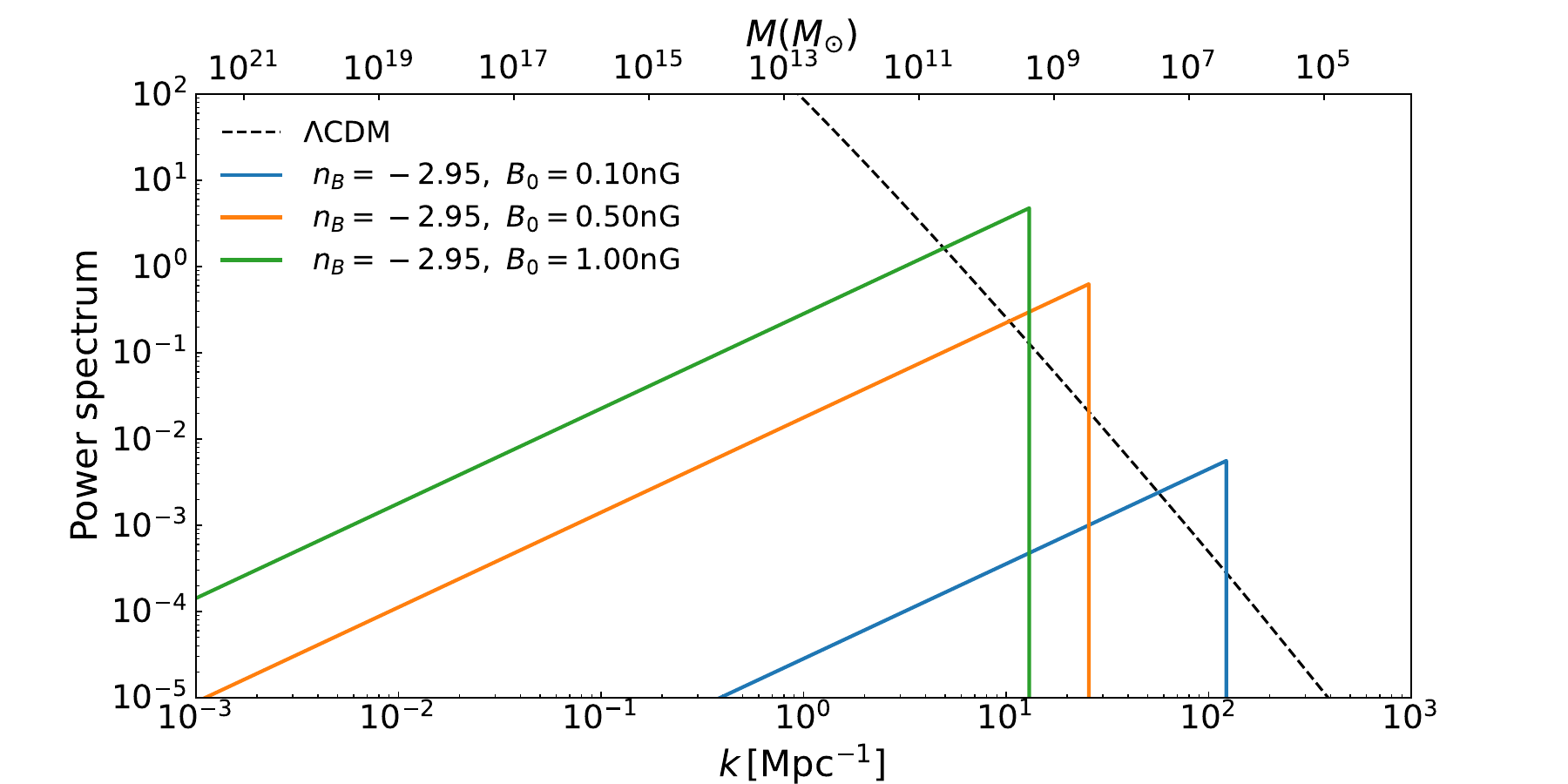}
\vspace{0.0cm}
\includegraphics[width=\columnwidth]{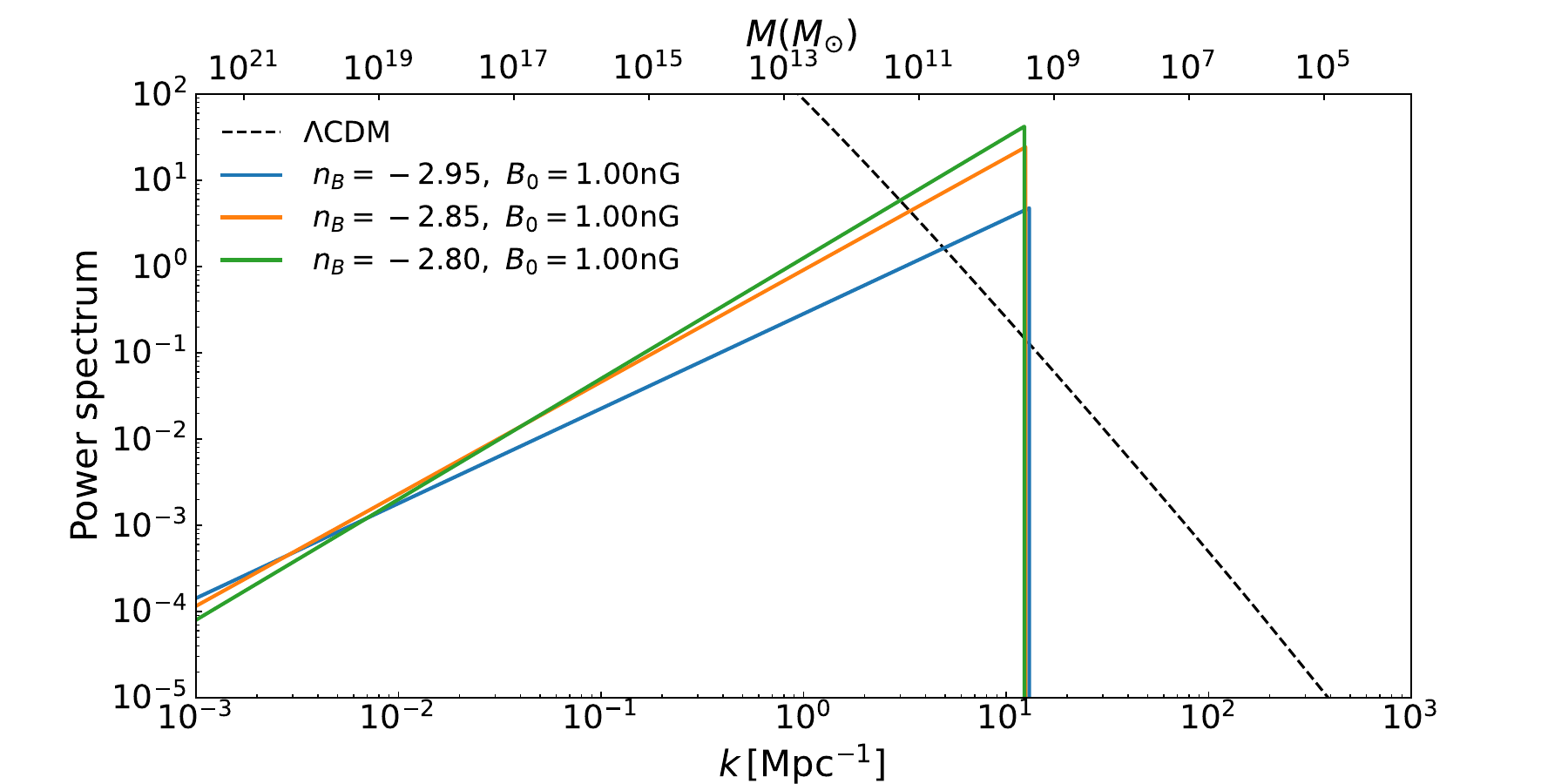}
\caption{The behaviour of the PMF-induced $\Pi(k)$ with different combinations of $B_0$ and $n_{\rm B}$ at today $z=0$.}
\label{PK_n_B}
\end{figure}

Here, we investigate the individual effects of $B_0$ and $n_B$ on the power spectrum $\Pi(k)$. As illustrated at the top of Figure \ref{PK_n_B}, it becomes apparent that $B_0$ predominantly serves to proportionally enhance the PMF-induced $\Pi(k)$ (displayed in logarithmic-logarithmic coordinates). Furthermore, the scale of the Jeans cut expands gradually with higher values of $B_0$. This observation underscores that excessively large values of $B_0$ would significantly boost the amplitude of the power spectrum, rendering it is incompatible with other Large-Scale Structure observational data \citet{Kahniashvili:2012dy}.
 
Conversely, the behavior of $n_B$ is more intricate, as depicted at the bottom of Figure \ref{PK_n_B}. As $n_{\rm B}$ approaches $-3$, the amplitude of $\Pi(k)$ is suppressed, accompanied by a subtle rotational effect. Interestingly, akin to $B_0$, the scale of the Jeans cut also exhibits a slight extension towards smaller scales.

\subsection{Impact of PMFs on SMD}
Besides the power spectrum, we also show the effects of $B_0$ and $n_B$ on the cumulative SMD in Figure \ref{SMD_without_loga_20}, which is directly related to the matter power spectrum through Eqs.(\ref{halomass}-\ref{rho_star}). 

For $B_0$, its influence on the SMD closely mirrors its effect on $\Pi(k)$, resulting in changes in the amplitude of $\rho_{*}$ represented by the dotted and dashed lines. Conversely, for $n_B$, illustrated by the solid line, the primary impact lies in modulating the slope of the SMD. When $n_{\rm B}$ approaches $-3$, the SMD curve exhibits a slight anticlockwise rotation. These distinctive effects offer valuable insights for our subsequent analysis.

In addition to these two PMF parameters, the SMD parameter $f_*^0$ also significantly affects the amplitude of $\rho_{*}$. As shown in Figure \ref{JWST}, higher values of $f_*^0$ correspond to elevated overall densities. $\rho_{*}$ plays a pivotal role in interpreting high-redshift data, with larger values aligning better with recent high-redshift observations.
Warm Dark Matter and Fuzzy Dark Matter can also be applied to SMD \citep{Gong_2023,2024RAA....24a5009L}.
But, their effect on the matter power spectrum is the opposite of that of PMFs, and in Gong et al. \citep{Gong_2023,2024RAA....24a5009L}, they interpret the SMD data by suppressing the matter power spectrum at high $k$ while raising $f_*^0$.
However, this alignment may introduce conflicts with other observational constraints and theoretical frameworks.

\begin{figure}
\centering
\includegraphics[width=\columnwidth]{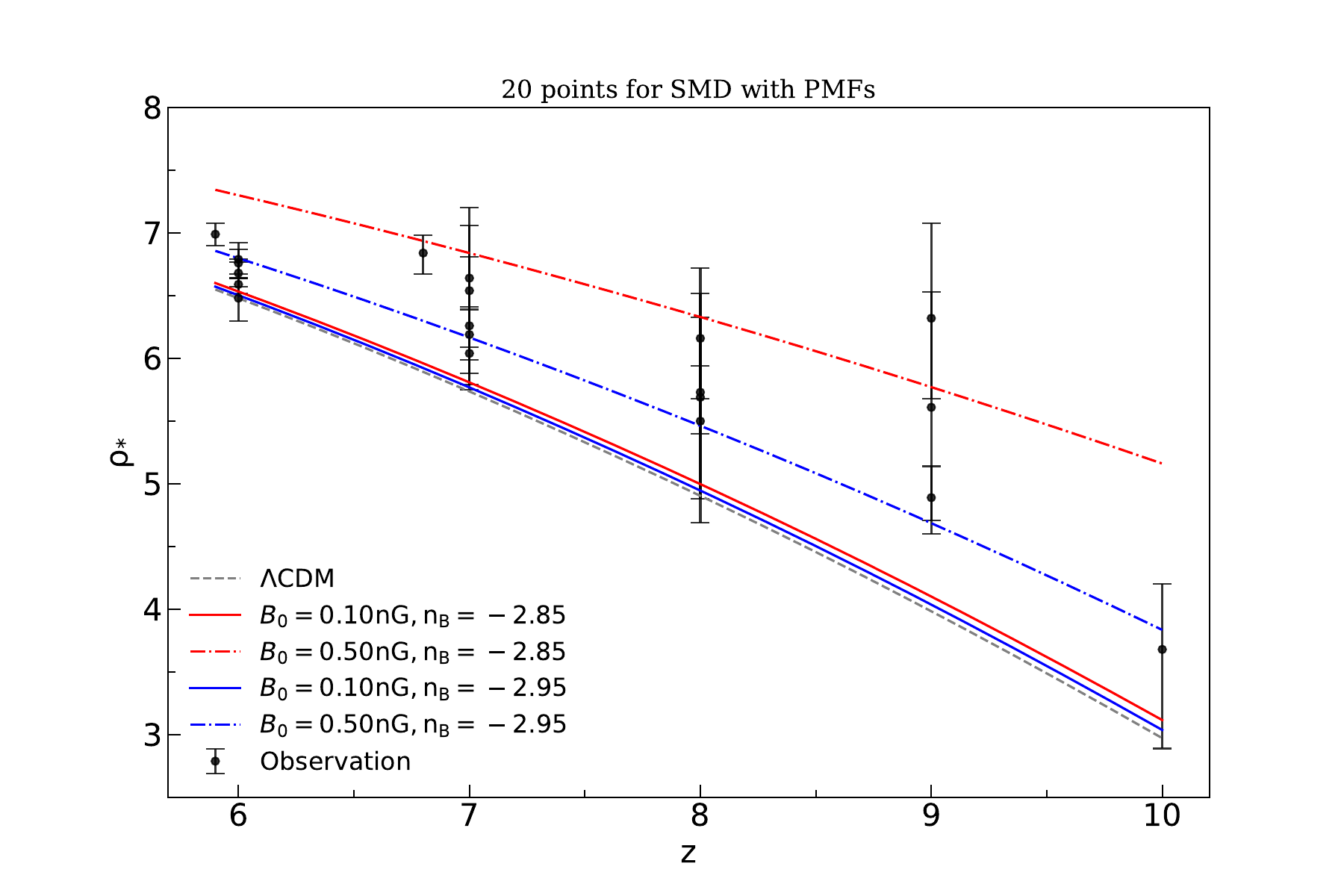}
\caption{The behaviour of the cumulative SMD with different combinations of $B_0$ and $n_{\rm B}$ with $f_*^0 = 0.1$. For comparison, we also plot the 20 SMD data (black) in the figure.}
\label{SMD_without_loga_20}
\end{figure}
%%%%%%%%%%%%%%%%%%%%%%%%%%%%%%%%%%%%%%%%%%%%%%%%%%

\subsection{14-points for $z>6.5$}

Before incorporating all 20 SMD data points, we firstly consider a conservative case by initially analyzing solely the last 14 SMD data points with $z > 6.5$, a redshift range where the double power law form of the SFE can be considered reliably applicable, thus we consider that $f_*^0$ in this interval is a fixed value that does not evolve with redshift. We list the constraints on three free parameters from 14 SMD data points in Table \ref{table_results}.

The amplitude parameter $B_0$ stands as a critical parameter in PMFs, directly correlating with the energy density of the magnetic field. A higher $B_0$ signifies a more potent magnetic field energy density, given by $\rho_{\mathrm{B}}(t)=B^2(t) /(8 \pi)$. Following the marginalization of other free parameters, we deduce a constraint of $B_0 < 4.44$ nG at a 95\% confidence level, a result akin to that derived from the Planck measurement \citep{PhysRevD.95.063506}. 

In our analysis, $n_B$ functions as a free parameter. Upon closer inspection of Figure \ref{14_20_points_f}, it becomes apparent that when $n_B$ approaches $-3$, the impact of the PMF-induced power spectrum becomes negligible. Consequently, regardless of how substantial $B_0$ is, both the power spectrum and the cumulative SMD remain largely unaltered. Conversely, when $n_B$ is significantly larger than $-3$, the evident departure in the power spectrum and SMD will not rule out large $B_0$ values, indicating a clear anti-correlation between $B_0$ and $n_B$. This results in the distribution of $B_0$ exhibiting a protracted tail, and the allowed area for $B_0$ expands significantly, leading to a weak constraint on $B_0$.

Regarding the spectrum index of PMFs $n_B$, most studies either do not treat it as a free parameter or subject it to multiple constraints. However, our analysis reveals that the cumulative SMD data can effectively constrain $n_B$, offering a significant result that distinguishes our study from previous works. 

In addition to the parameters $B_0$ and $n_B$ related to PMFs, we also obtained the constraint on the SFE parameter: $f_*^0 = 0.123 \pm 0.042$ at the 68\% confidence level. This result aligns well with predictions from theoretical models of the star formation process, which typically suggest $f_*^0 \sim 0.1$. This consistency reinforces the validity of our model and analysis.

\begin{table}
\centering
\caption{The 68\% and 95\% confidence level of upper bounds on the PMFs amplitude $B_0$ (nG) and the spectral index $n_B$, as well as the median value and the 68\% confidence level of the SMD parameter $f_*^0$, in different cases.}
\renewcommand{\arraystretch}{2.5}
\scriptsize
\tabcolsep=0.07cm
\begin{tabular}{l|c|c|c|c|c}
\hline \hline 
& $B_0 (68\%)$ & $B_0 (95\%)$ & $n_B(68\%)$ &$n_B(95\%)$ & $f_*^0(68\%)$\\
\hline
14-points& $<1.05$ & $<4.44$ & $<-2.79$ & $<-2.24$ & $0.123\pm0.042$\\
20-points& $<1.70$ & $<4.20$ & $<-2.69$ & $<-2.10$ & $0.044\pm0.012$\\
$n_B=-2.85$& $<0.352$ & $<1.33$ & $-$ &$-$ & $0.146 \pm 0.071$\\
$n_B=-2.90$& $<0.465$ & $<2.21$ & $-$ &$-$ & $0.145\pm0.090$\\
$n_B=-2.95$& $<0.979$ & $<3.90$ & $-$ &$-$ & $0.142\pm0.095$\\
\hline \hline
\end{tabular}
\label{table_results}
\end{table}

%%%%%%%%%%%%%%%%%%%%%%%%%%%%%%%%%%%%%%%%%%%%%%%%%%
\subsection{20-points for $z>5.5$}

In the previous subsection, we conservatively used 14 SMD data points with $z > 6.5$. The remaining 6 data points are around $z \approx 6$, where the double power law form of the SFE could still be roughly reliable. Therefore, we now incorporate all 20 SMD data points into our analysis to provide a more comprehensive constraint on the parameters, which listed in Table \ref{table_results}.

The most notable change is observed in the constraint on $f_*^0$, where the optimal value of $f_*^0$ shifts to 0.044, markedly departing from the theoretical expected value of 0.1 derived from theoretical models of the star formation process. This shift primarily arises due to the inclusion of six new data points, which notably fall below the theoretical expectations for the SMD. Given that the parameter $f_*^0$ is highly influenced by the magnitude of the SMD, the permissible range of $f_*^0$ values is significantly suppressed by these SMD data.

Therefore, we need to increase the amplitude of PMFs $B_0$ to counterbalance the impact of the decrease in $f_*^0$ on the cumulative SMD. Consequently, our measurement results for $B_0$ are slightly weaker compared to those when using the 14 data points. The upper limit of the 68\% confidence level increases from 1.05 nG to 1.70 nG, while the upper limit of the 95\% confidence level remains almost unchanged. In the meanwhile, the constraints of $n_B$ will undergo corresponding changes. The upper limits of the 68\% and 95\% confidence levels both increase to varying degrees, as clearly shown in Figure \ref{14_20_points_f}.

Due to the inconsistency with the theoretical expectations of $f_*^0$ concerning the SMD, for the subsequent analysis, we will continue to utilize the 14 SMD data points to derive the final constraints on the relevant parameters.

\begin{figure}
\centering
\includegraphics[width=\columnwidth]{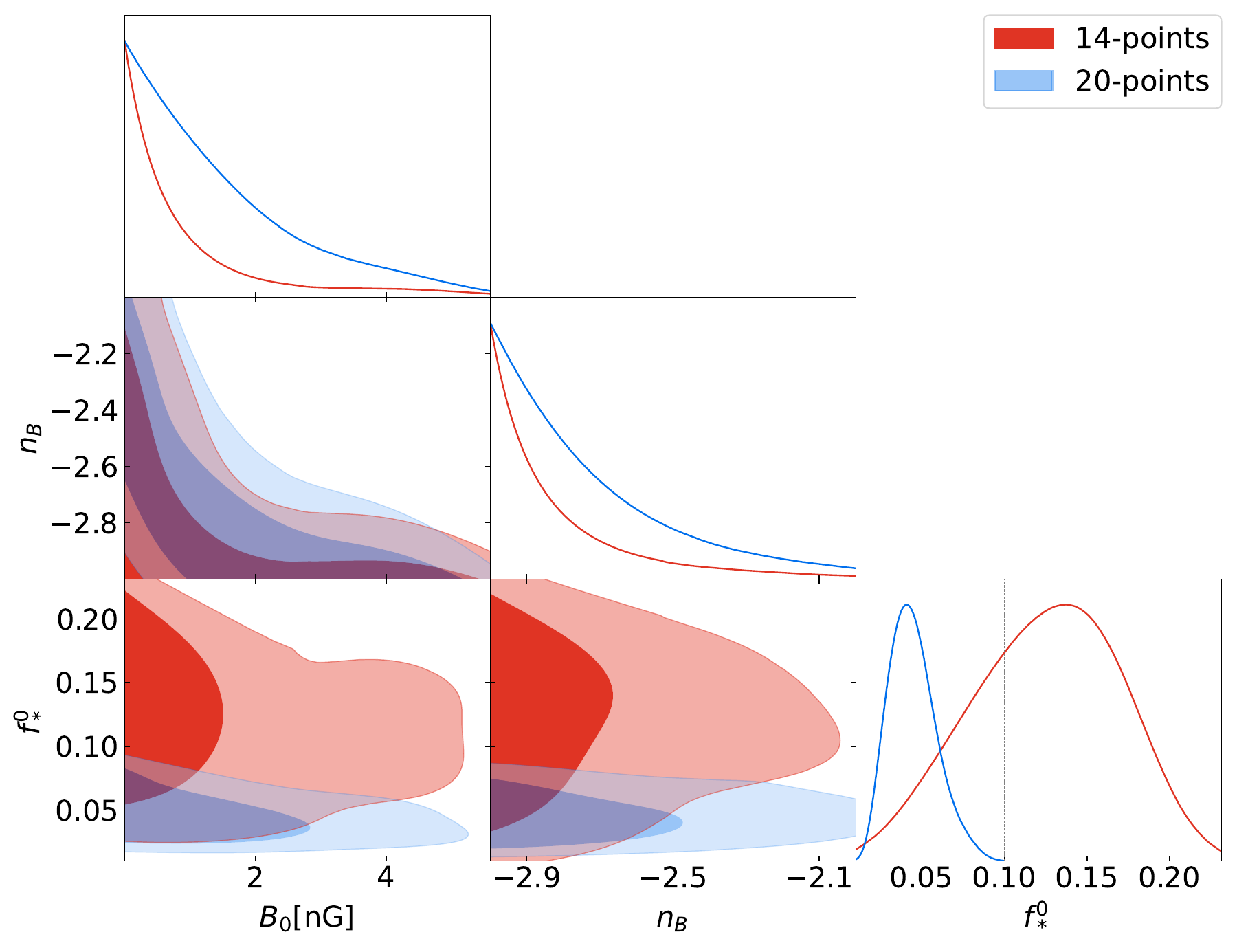}
\caption{1D and 2D dimensional constraints on parameters of $B_0$, $n_B$ and $f_*^0$ from the 14 and 20 SMD data points, respectively. The vertical gray dotted line here represents $f_*^0=0.1$. The 2D-dimensional contours encompass 68\% and 95\% of the probability.}
\label{14_20_points_f}
\end{figure}

%%%%%%%%%%%%%%%%%%%%%%%%%%%%%%%%%%%%%%%%%%%%%%%%%%
\subsection{Fixed the spectrum index}

In previous studies, constraints on the spectrum index $n_B$ have often been weak, leading to situations where $n_B$ is fixed at certain values to discuss the constraints on the amplitude $B_0$ \citep{pandey2015reionization,PhysRevLett.123.021301,2013ApJ...762...15P,sanati2024dwarf}. 
As mentioned earlier, there is a clear correlation between $n_B$ and $B_0$, such that once $n_B$ is constrained to specific values, the constraints on $B_0$ become significantly stronger. 

Therefore, in our study, it is essential to explore the constraints on $B_0$ while setting $n_B$ to some typical values, such as $-2.95$, $-2.9$, and $-2.85$, as utilized in previous studies like \citet{pandey2015reionization}.
It is worth mentioning that, the specific value of $n_B$ is not clearly given by the model such as PMFs generated during inflation. According to the work of previous researchers, we have widely selected the interval of $n_B$ from $-3$ to $-2.5$, so our values for $n_B$ are reasonable generated during inflation model.
In these cases, two free parameters, $B_0$ and $f_*^0$ are considered. The corresponding constraint results from the 14 SMD data points are presented Table \ref{table_results} and Figure \ref{14_f_fixed}.

Firstly, we observe that in these three cases, the constraints on the SFE parameter $f_*^0$ exhibit a notable similarity. The marginalized distributions are considerably broadened, with the maximum likelihood values shifting from 0.123 to approximately 0.2. Since $f_*^0$ and $B_0$ both affect the amplitude of the cumulative SMD, a larger value of $f_*^0$ will lead to a tighter constraint on $B_0$. Therefore, we can see that the constraints on $B_0$ are now quite stringent, when compared with the previous results.

As we mentioned before, there is a anti-correlation between $B_0$ and $n_B$. The further $n_B$ deviates from $-3$, the stronger the constraints on $B_0$ become from the SMD data. Therefore, when we fix the spectrum index to be $n_B=-2.95,\,2.90,\,2.85$, the 95\% C.L. upper limits of the amplitude are $B_0<3.90$ nG, $B_0<2.21$ nG and $B_0<1.33$ nG, respectively. These constraints are a little bit weaker than those from the reionization history of Universe \citep{pandey2015reionization}, but is comparable with the constraint from the CMB measurements \citep{PhysRevD.95.063506}.

\begin{figure}
\centering
\includegraphics[width=1\columnwidth]{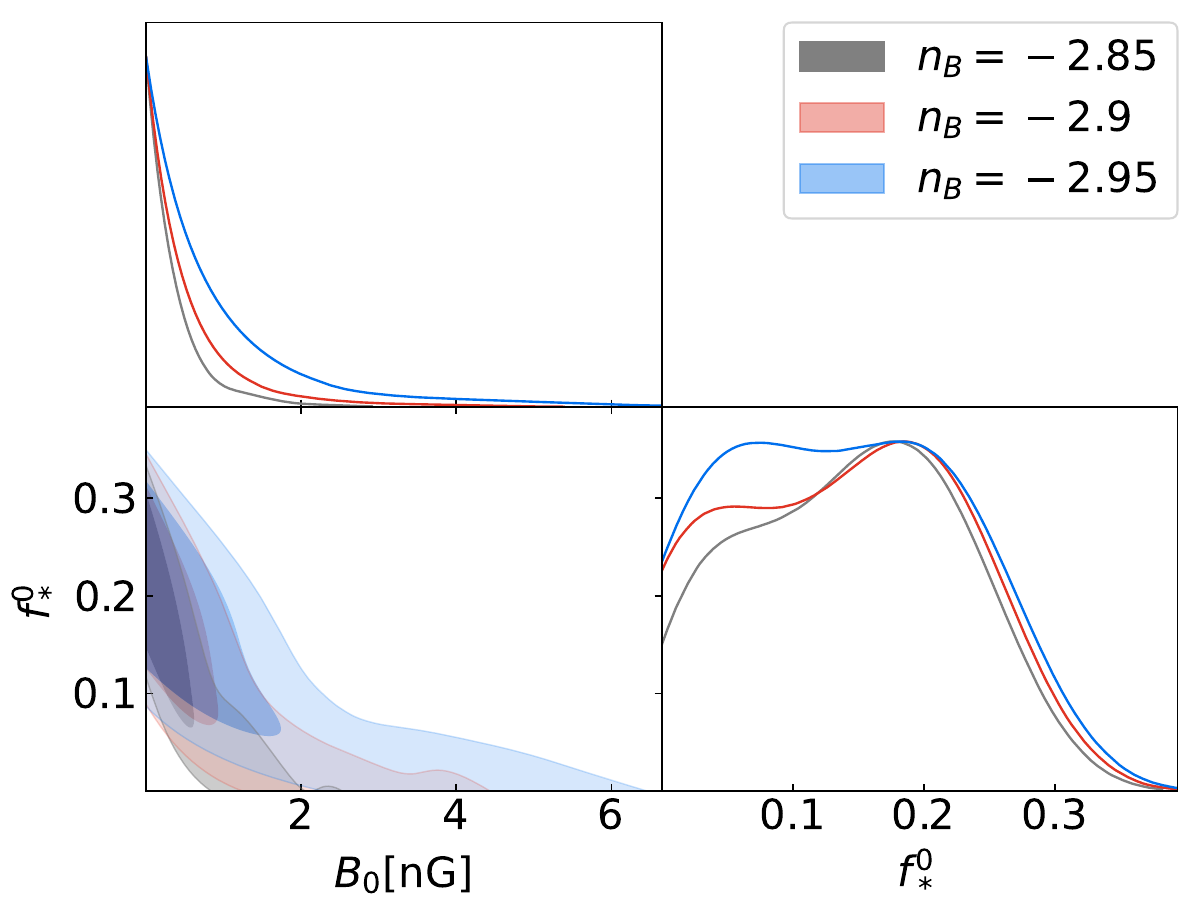}
\caption{Constraints on $f_*^0$ and $B_0$ from the 14 SMD data points, when fix the spectrum index as $n_B=-2.95,-2.9,-2.85$. The 2D-dimensional contours encompass 68\% and 95\% of the probability.}
\label{14_f_fixed}
\end{figure}

%%%%%%%%%%%%%%%%%%%%%%%%%%%%%%%%%%%%%%%%%%%%%%%%%%
\subsection{Measurements from JWST}

Finally, we integrate observations from the James Webb Space Telescope (JWST) into our analysis. Early release observations from the JWST have revealed galaxies with stellar masses reaching approximately $10^{11}M_{\odot}$ out to $z>6$ \citep{2023Natur.616..266L}. These galaxies are estimated to have formed roughly 1 billion years after the Big Bang. The existence of such massive galaxies at such early cosmic epochs poses a significant challenge to the standard cosmological model, $\Lambda \mathrm{CDM}$, and prevailing star formation theories \citep{PhysRevLett.132.061002}. These high-redshift data significantly surpass previous observations, underscoring the need for a more comprehensive understanding.

However, it is important to note that the JWST relies on current photometric redshift (photo-$z$) estimation methods, which can introduce significant uncertainties. Studies such as \citet{Gong_2023} suggest that different spectral energy distribution (SED) templates and photo-$z$ programs can yield varying redshift estimates.

Regardless of the uncertainties in high-redshift measurements, the star formation efficiency $f_*$ is another important issue related to the JWST result. As we know, $f_*$ can vary based on redshift, star formation theory, galaxy formation theory, and the cosmological model. While $f_*^0$ values are typically considered to be less than 0.1 based on lower redshift observations \citep{2022ApJ...938L...5M,2023NatAs...7..731B}, higher values, even approaching $f_*^0 \sim 1$, are theoretically plausible at higher redshifts. 

Recently, several studies indicate that it is impossible to maintain $f_*^0$ at 0.1 to interpret JWST observations under the $\Lambda \mathrm{CDM}$ model. 
For instance, \citet{2023JCAP...10..012F} uses $Planck$ large-scale polarization data and finds that there is a significant discrepancy between the full Planck CMB angular spectrum measurements and the JWST observations if $f_*^0<0.3$.
They indicates that a value of $f_*^0$ greater than 0.3 is needed for compatibility. When $f_*^0$ is 0.32, the discrepancy decreases to less than 3 $\sigma$, as also seen in Figure 2 of \citet{2023NatAs...7..731B}. However, as mentioned in \citet{Gong_2023}, a high $f_*^0$ would conflict with the cosmic reionization history.

In our study, we present a comparative analysis between theoretical calculations and observational data, incorporating PMFs, as depicted in Figure \ref{JWST}. We first set $f_*^0 = 0.1$ to explore whether maintaining $f_*^0$ at this value, even at high redshifts, can account for the JWST data. The solid black line in the upper panel of Figure \ref{JWST} represents the $\Lambda$CDM model with $f_*^0 = 0.1$. However, the observed JWST data (blue points) significantly exceeds theoretical predictions, even when $f_*^0$ is increased to 0.3 (black line in the lower panel of Figure \ref{JWST}). When we take the PMFs into account, we find that for $f_*^0 = 0.1$, agreement with observational data is achievable, although it requires substantial values of $ B_0 $ (minimum 4 nG) and $ n_B$ ($ < -2.90 $), as shown by the dotted purple and solid blue lines.

Considering that these values of $ B_0 $ and $n_B$ may already be ruled out by other measurements, we utilize larger values of $f_*^0$, such as 0.3 or 0.5, in the calculations. We find that in these cases, large values of $ B_0 $ and $n_B$ are not necessary. Typical values of $B_0 = 0.5$ and $n_B = -2.9$ are sufficient to explain the JWST results, as illustrated in the lower panel of Figure \ref{JWST}.

Based on this calculation, we conform that PMFs can play a similar role with $f_*^0$ in enhancing the cumulative SMD. Therefore, it is possible to increase $\rho_{*}$ to explain the observed data from JWST without increasing $f_*^0$ and affecting the cosmic reionization history. 

\begin{figure}
\centering
\includegraphics[width=\columnwidth]{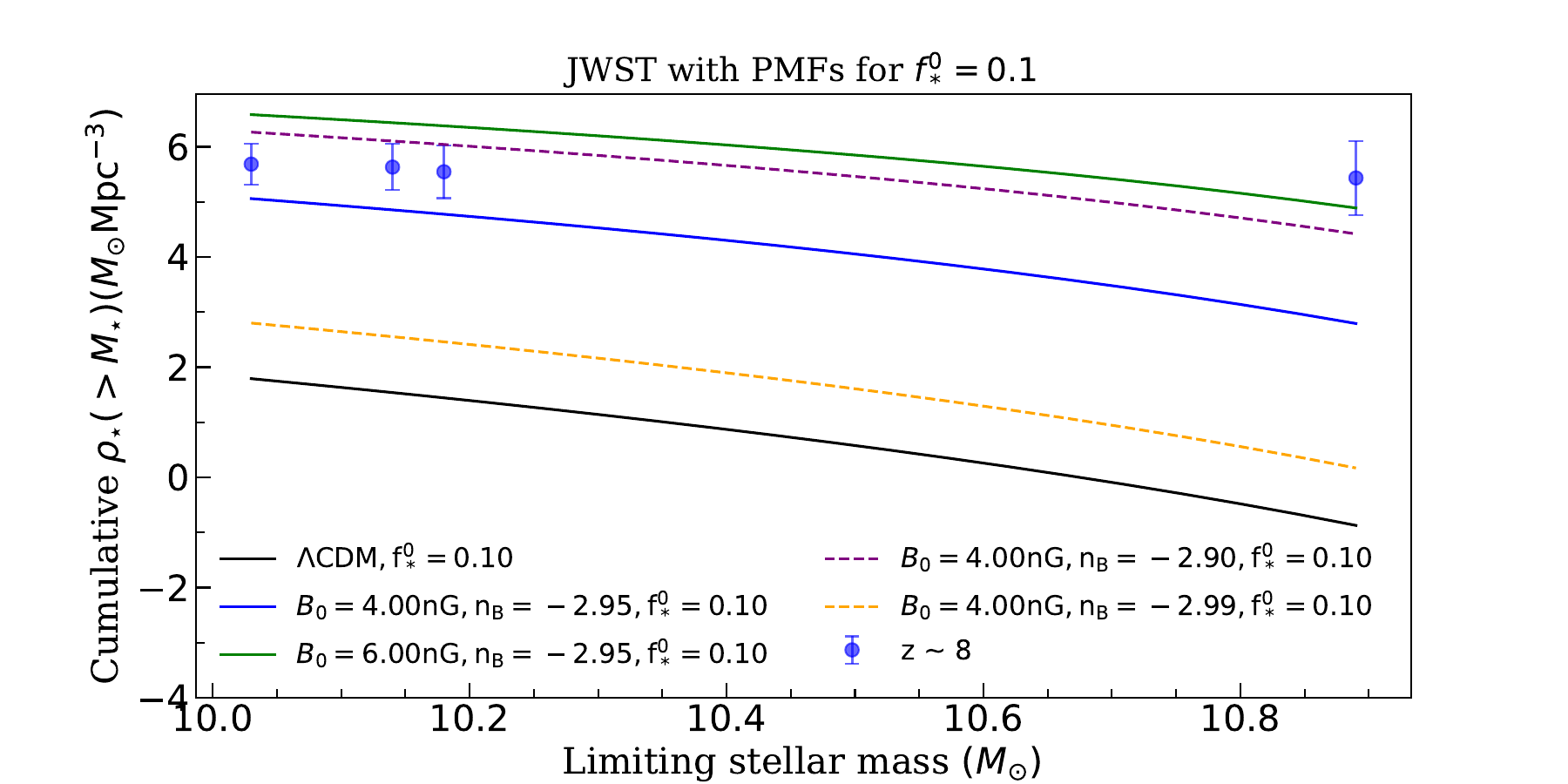}
\vspace{0.0cm}
\includegraphics[width=\columnwidth]{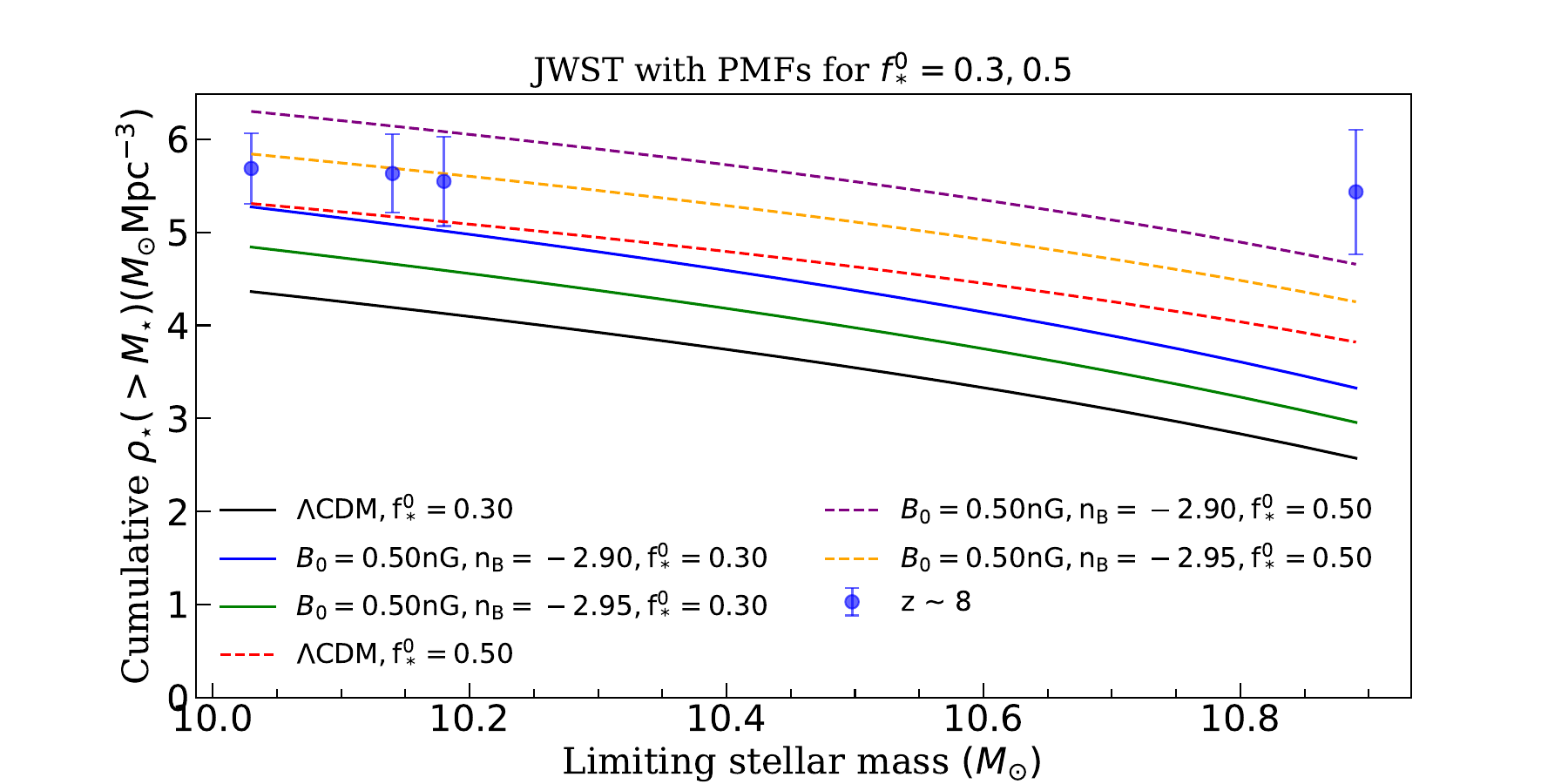}
\caption{The cumulative SMD results in different combinations of $n_B$ and $B_0$, with $f_*^0=0.1$ (upper panel) and $f_*^0=0.3,\,0.5$ (lower panel). We also plot the JWST data point at $z\sim 8.0$ (blue) for comparison.}
\label{JWST}
\end{figure}
%%%%%%%%%%%%%%%%%%%%%%%%%%%%%%%%%%%%%%%%%%%%%%%%%%
\section{Conclusion}\label{section:6}

The study of small-scale phenomena in the universe, as revealed by the observation of the cumulative SMD, provides valuable insights that complement other cosmological probes like the CMB and BAO, which predominantly address large to medium scales of the universe. In this research, we have gathered 20 data points from cumulative SMD observations and examined their potential to constrain the parameters of PMFs. Our main conclusions are summarized as follows:

Through the close relationship between SMD and matter power spectrum, we firstly investigate how the PMF parameters, such as $B_0$ and $n_B$, influence the power spectrum $P(k)$ and the SMD, respectively. We find that the $B_0$ affects the amplitude of the SMD, while the SMD curve exhibits a slight rotation $n_B$ approaches $-3$.

Subsequently, we analyze 14 SMD data points with redshifts $z > 6.5$ to derive upper limits on the parameters $B_0$ and $n_B$, finding $B_0 < 4.44$ nG and $n_B < -2.24$ at the 95\% confidence level. These constraints align with previous results obtained from CMB measurements. Additionally, this analysis enables us to constrain the star formation efficiency parameter, yielding $f^*_0 = 0.123 \pm 0.042$ (68\% C.L.), consistent with theoretical models of the star formation process.

We also extend our analysis to include lower redshifts by incorporating the remaining six data points at $z \sim 6$. When utilizing all 20 SMD data points to constrain the PMF model, the most notable change is observed in the constraint on the SFE parameter $f^*_0$, where the optimal value shifts to 0.044, significantly deviating from the theoretically expected value of 0.1. Consequently, the constraints on $B_0$ and $n_B$ are slightly weaker compared to those derived from the 14 high-redshift data points alone, with $B_0 < 4.20$ nG and $n_B < -2.10$ at the 95\% confidence level.

Our results indicate a clear correlation between $n_B$ and $B_0$. By fixing $n_B$ at specific values, such as $-2.95$, $-2.9$, and $-2.85$, we obtain significantly stronger constraints on $B_0$. Specifically, the 95\% C.L. upper limits on $B_0$ are $<3.90$ nG, $<2.21$ nG, and $<1.33$ nG, respectively. Concurrently, to align with the data points, the star formation efficiency parameter $f^*_0$ adjusts from 0.123 to approximately 0.2.

Finally, we explore the potential of PMFs to account for observations from the JWST, which show significant deviations from the $\Lambda$CDM model. By setting $f^*_0=0.1$, we find that agreement with observational data can be achieved, though it necessitates substantial values for both $B_0$ and $n_B$. If we further explore scenarios with larger values of $f^*_0$, such as 0.3 or 0.5, we find that typical values of $B_0 = 0.5$ nG and $n_B = -2.9$ are sufficient to explain the JWST observations. Our findings suggest that by selecting appropriate PMF parameters, it is possible to account for these observations without significantly increasing the star formation efficiency.

Given the current scarcity of high-redshift observations compared to other types of observational data, as well as the significant statistical and systematic errors, future efforts should prioritize acquiring more accurate and comprehensive data. We anticipate that understanding the impact of PMFs on star and galaxy formation at small scales could eventually aid in interpreting early observations from instruments like the JWST \citep{2023Natur.616..266L}. This could open new avenues for exploring the early universe and enhancing our understanding of cosmological phenomena.

%%%%%%%%%%%%%%%%%%%%%%%%%%%%%%%%%%%%%%%%%%%%%%%%%%
\section*{Acknowledgements}
This work is supported by the National Nature Science Foundation of China under grant Nos. U1931202 and 12021003, the National Key R\&D Program of China No. 2020YFC2201603, and the Fundamental Research Funds for the Central Universities.

\bibliography{Reference}
\bibliographystyle{aasjournal}

\end{document}